\documentclass[preprint,preprintnumbers,amsmath,amssymb]{revtex4}
\usepackage{graphicx}

\begin{document}

\bibliographystyle{pnas}

\title{Scaling laws of human interaction activity}

\author{ Diego Rybski$^{1}$, Sergey V. Buldyrev$^{2}$,
  Shlomo Havlin$^{3}$,\\
  Fredrik Liljeros$^{4}$, and Hern\'an A. Makse$^{1}$}

\affiliation{ $^1$Levich Institute and Physics Department, City
  College of New York,
  New York, NY 10031, USA\\
  $^2$Department of Physics, Yeshiva University,
  New York, NY 10033, USA\\
  $^3$Minerva Center and Department of Physics, Bar-Ilan University,
  Ramat-Gan 52900, Israel\\
  $^4$Department of Sociology, Stockholm University, S-10691
  Stockholm, Sweden }

\date{\today}

\begin{abstract}
  Even though people in our contemporary, technological society are
  depending on communication, our understanding of the underlying laws
  of human communicational behavior continues to be poorly understood.
  Here we investigate the communication patterns in two social
  Internet communities in search of statistical laws in human
  interaction activity. This research reveals that human communication
  networks dynamically follow scaling laws that may also explain the
  observed trends in economic growth. Specifically, we identify a
  generalized version of Gibrat's law of social activity expressed as
  a scaling law between the fluctuations in the number of messages
  sent by members and their level of activity. Gibrat's law has been
  essential in understanding economic growth patterns, yet without an
  underlying general principle for its origin. 
  We attribute this scaling law to long-term correlation 
  patterns in human activity, which surprisingly span from days to the 
  entire period of the available data of more than one year. 
  Further, we provide a mathematical framework that relates the 
  generalized version of Gibrat's law to the long-term 
  correlated dynamics, which suggests that the same underlying mechanism 
  could be the source of Gibrat's law in economics, ranging from large 
  firms, research and development expenditures, gross domestic product of 
  countries, to city population growth.  
  These findings are also of importance for
  designing communication networks and for the understanding of the
  dynamics of social systems in which communication plays a role, such
  as economic markets and political systems.
\end{abstract}

\maketitle

\section{Introduction}

The question of whether unforeseen outcomes of social activity follow
 emergent statistical laws has been an acknowledged problem in the
 social sciences since at least the last decade of the 19th century
 \cite{MertonR1936,WeberM1968,GiddensA1993,DurkheimE1997}.
Earlier discoveries include Pareto's law for income distributions
 \cite{ParetoV1896}, Zipf's law initially applied to word frequency
 in texts and later extended to firms, cities and others
 \cite{ZipfG1932}, and Gibrat's law of proportionate growth in
 economics \cite{GibratR1931,SuttonJ1997,GabaixX1999}.

Social networks are permanently evolving and Internet communities 
are growing each day more.
Having access to the communication patterns of
Internet users opens the possibility to unveil the origins of
statistical laws that lead us to the better understanding of human
behavior as a whole.
In this paper, we analyze the dynamics of sending messages in two
Internet communities in search of statistical laws of human
communication activity.
The first online community (OC1)
is mainly used by the group of men who have sex with men (MSM)
\footnotemark \footnotetext{The study of the de-identified MSM dating
  site network data was approved by the Regional Ethical Review board
  in Stockholm, record 2005/5:3.}.  The data consists of over
$80,000$~members and more than $12.5$~million messages sent during
$63$~days.
The target group of the second online community (OC2)
is teenagers \cite{HolmeEL2004}.  The data covers 492 days of activity
with more than 500,000 messages sent among almost 30,000 members.
Both web-sites are also used for social interaction in general.  All
data are completely anonymous, lack any message content and consist
only of the time when the messages are sent and identification numbers
of the senders and receivers.

The act of writing and sending messages is an example of an
intentional social action.  In contrast to routinized behavior,
the actants are aware of the purpose of their actions
\cite{WeberM1968,GiddensA1993}.  Nevertheless, the emergent properties
of the collective behavior of the actants are unintended.  In
Fig.~\ref{fig:example}a we show a typical example of the activity of a
member of OC1 depicting the times when the member sends messages.
Figure~\ref{fig:example}b provides the cumulative number of messages
sent (green curve) compared with a random surrogate data set (brown
curve) obtained by shuffling the data, as discussed below.  As would
be expected, there are large fluctuations in the members' activity
when compared with a random signal
\cite{PaxsonF1995,DewesWF2003,BarabasiAL2005,ZhouKKWH2008}.
The messages sent at random display small temporal fluctuations while
the OC1 member sends many more messages in the beginning and much less
at the end of the period of data acquisition (as also seen in
Fig.~\ref{fig:example}c, displaying the number of messages sent per
day).  While such extreme events or bursts have been documented for
many systems, including e-mail and letter post communication, 
instant messaging, web browsing and movie watching
\cite{PaxsonF1995,DewesWF2003,BarabasiAL2005,OliveiraB2005,ZhouKKWH2008}, 
their origin is still an open question.

\section{Results}

\section*{Growth in the number of messages}

The cumulative number, $m_j(t)$, expresses how many messages have been
sent by a certain member~$j$ up to a given time~$t$ [for a better
readability we will not write the index~$j$ explicitly, $m(t)$, see
details on the notation in the Supporting Information (SI)
Sec.~I]. 
The dynamics of $m(t)$ between times $t_0$
and $t_1$ within the period of data acquisition $T$ ($t_0<t_1\le T$)
can be considered as a growth process, where each member exhibits a
specific growth rate~$r_j$ ($r$~for short notation):
\begin{equation}
r\equiv\ln\frac{m_1}{m_0}
\, ,
\label{eq:rlnm1m0}
\end{equation}
where $m_0\equiv m(t_0)$ and $m_1\equiv m(t_1)$ are the number of
messages sent until $t_0$ and $t_1$, respectively, by every member.
To characterize the dynamics of the activity, we consider two
measures. {\it (i)} The conditional average growth rate, $\langle
r(m_0)\rangle$, quantifies the average growth of the number of
messages sent by the members between~$t_0$ and~$t_1$ depending on the
initial number of messages, $m_0$.  In other words, we consider the
average growth rate of only those members that have sent~$m_0$
messages until~$t_0$ (see Methods, Sec.~\ref{sec:matmathe} for more
details).  {\it (ii)} The conditional standard deviation of the growth
rate for those members that have sent~$m_0$ messages until~$t_0$,
$\sigma(m_0)\equiv\sqrt{\langle(r(m_0)-\langle
  r(m_0)\rangle)^2\rangle}$, expresses the statistical spread or
fluctuation of growth among the members depending on $m_0$.  Both
quantities are relevant in the context of Gibrat's law in economics
\cite{GibratR1931,SuttonJ1997,GabaixX1999}
which proposes a proportionate growth process entailing the assumption
that the average and the standard deviation of the growth rate of a
given economic indicator are constant and independent of the specific
indicator value. That is, both $\langle r(m_0)\rangle$ and
$\sigma(m_0)$ are independent of
$m_0$ \cite{GabaixX1999}

In Fig.~\ref{fig:gibrat}a,b we show the results of $\langle
r(m_0)\rangle$ and $\sigma(m_0)$ versus~$m_0$ for both online
communities.  We find that the conditional average growth rate is
fairly independent of $m_0$. On the other hand, the standard deviation
decreases as a power-law of the form:
\begin{equation}
\sigma(m_0)\sim m_0^{-\beta}
\, .
\label{eq:srm0simm0b}
\end{equation}
We obtain by least square fitting the exponents $\beta_{\rm
  OC1}=0.22\pm 0.01$ for OC1 and $\beta_{\rm OC2}=0.17\pm 0.03$ for
OC2 (the values deviate slightly for large~$m_0$ due to low
statistics).  Although the web-sites are used by different member
populations, the power-law and the obtained exponents are quite
similar.  The exponents are also close to those reported for growth in
economic systems such as firms and countries ($0.15-0.18$,
\cite{StanleyABHLMSS1996}), research and development expenditures at
universities ($0.25$, \cite{PlerouAGMS1999}),
scientific output ($0.28-0.4$,
\cite{MatiaALMS2005}), 
and city population growth ($0.19-0.27$,
\cite{RozenfeldRABSM2008}).  The approximate agreement between the
exponents obtained for very different systems (social or of human
origin) can be considered as a generalization of Gibrat's law,
suggesting that the mechanisms behind the growth properties in
different systems may originate in the human activity represented by
Eq.~(\ref{eq:srm0simm0b}).

Figures~\ref{fig:gibrat}c and d depict the results when we randomize
the data of OC1 and OC2, respectively (see Sec.~\ref{sec:matmathe}
for details of the randomization procedure), such that any temporal
correlations are removed.  The typical dynamics for such surrogate
data set are shown in Fig.~\ref{fig:example}b (the brown curve) displaying
a clear random pattern of small fluctuations in comparison with the
original data of larger fluctuations (green curve).  We find that the
random signal displays a close to constant average growth rate $\langle
r(m_0)\rangle$ and that the fluctuations
behave as in Eq.~(\ref{eq:srm0simm0b}) but with an exponent
$\beta_{\rm rnd}=1/2$ (Fig.~\ref{fig:gibrat}c,d).  The origin of this
value has a simple explanation: If an isolated individual randomly
flips an ideal coin with no memory of the previous attempt, then the
fluctuations from the expected value of the fraction of obtained heads
decay as a square-root of the number of throws, implying $\beta_{\rm
  rnd}=1/2$. In contrast to randomness, here we hypothesize that the
origin of the generalized version of Gibrat's law with $\beta<1/2$ in
Eq.~(\ref{eq:srm0simm0b}) is a non-trivial long-term correlation in
communication activity. These correlations possibly arise from
internal and external stimuli from other members transmitted through
the highly connected network of individuals, an effect that is absent
in the randomized data.
The exponent value of $\beta\approx 0.2$ for OC1 and OC2 implies
that the fluctuations of very active members are smaller than the ones
of less active members, but they are significantly larger compared to
the random case (compare Fig.~\ref{fig:gibrat}a,b with
Fig.~\ref{fig:gibrat}c,d).

\section*{Long-term correlations}

The exceptional quality of the data (more than 10 million messages
spanning several effective decades of magnitude in terms of both
activity and time) allows to test the above hypothesis by
investigating the presence of temporal correlations in the
individuals' activity.
We aggregate the data to records of messages per day (an example is shown 
in Fig.~\ref{fig:example}c) to avoid the daily cycle in the activity and 
analyze the number of messages sent by individuals per day, $\mu(t)$,
where $t$ denotes the day [$m(t)\equiv \sum_{t'=1}^{t}\mu(t')$,
Figs.~\ref{fig:example}d-f show the color coded daily activity of
three members in OC1].  For every member we obtain a record of a
length of $63$~days (OC1) or $492$~days (OC2).  We note that former
studies reporting Eq.~(\ref{eq:srm0simm0b}) such as 
\cite{StanleyABHLMSS1996,PlerouAGMS1999,MatiaALMS2005,RozenfeldRABSM2008}
typically were not based on data with temporal resolution as we use it
here, and therefore were not able to investigate its origin in terms
of temporal correlations.

We quantify the temporal correlations in the members' activity by
mapping the problem to a one-dimensional random walk.  The quantity
$Y(t)\equiv\sum_{t'=1}^{t}\left(\mu(t')-\langle\mu(t)\rangle\right)$,
where $\langle \mu(t)\rangle$ is the average of the corresponding
record $\mu(t)$, represents the position of the random walker that
performs an up or down step given by $\mu(t')-\langle\mu(t)\rangle$ at
time step~$t'$. The correlations after $\Delta t$ steps are reflected
in the behavior of the root-mean-square displacement $F(\Delta
t)\equiv \sqrt{\langle\left[Y(t+\Delta t)-Y(t)\right]^2\rangle}$ 
\cite{Feder1988}, where $\langle\cdot\rangle$ is the average over~$t$ 
and members. 
If the activity~$\mu(t)$ is {\it uncorrelated} or
{\it short-term correlated}, then one obtains $F(\Delta t)\sim (\Delta
t)^{1/2}$, Fick's law of diffusion, after some cross-over time.  In
the case of {\it long-term correlations}, the result is a power-law
increase
\begin{equation}
F(\Delta t)\sim (\Delta t)^H
\, , 
\label{eq:fdtsimdth}
\end{equation}
where $H>1/2$ is the fluctuation exponent (also known as Hurst
exponent \cite{Feder1988}).  In statistical physics, long-term
correlation or persistence is also referred to as long-term
``memory''.  Since, in general, the records might be affected by
trends, we use the standard Detrended Fluctuation Analysis (DFA)
\cite{PengBHSSG94} to 
calculate~$H$ (see SI Sec.~III for a detailed description).

The results for OC1 are shown in Figs.~\ref{fig:dfa}a,b, where we
calculate Eq.~(\ref{eq:fdtsimdth}) by separating the members in groups
with different total number of messages sent by the members, $M$.  We
find that $F(\Delta t)$ asymptotically follows a power-law with
$H\approx 1/2$ for the less active members who sent less than $10$
messages in the entire period ($M<10$). The dynamics of the more
active members display clear long-term correlations. We find that the
fluctuation exponent increases to $H\approx 0.75$ for members with
$M>10^3$ (see Fig.~\ref{fig:dfa}b).  The smaller value of $H$ for less
active members could be due to the small amount of information that
these members provide in the available time of data acquisition.  When
we shuffle the data to remove any temporal correlations, we obtain the
random exponent $H_{\rm rnd}= 1/2$ (as seen in Fig.~\ref{fig:dfa}b),
confirming that the correlations in the data are due to temporal
structure.

The dynamics of the message activity in OC2 is similar to OC1 (see
Fig.~\ref{fig:dfa}c).
On large time scales we measure the fluctuation exponent increasing
from~$H\approx 1/2$ to~$H\approx 0.9$ with increasing $M$ (the
exponents for very active members are based on poor statistics and
therefore carry large error bars).  Analogous to the results obtained
for OC1, there are no correlations in the shuffled records 
($H_{\rm rnd}=1/2$ in Fig.~\ref{fig:dfa}d).
The fact that $H>1/2$ means that a sudden burst in activity of a
member persists on times scales ranging from days to years. The
distribution of activity is self-similar over time.
Similar correlation results have been found in traded values of stocks
and email data \cite{EislerBK2008}.

\section*{Relation between $\beta$ and $H$}

Next, we elaborate the mathematical framework that relates the growth
process Eq.~(\ref{eq:srm0simm0b}) to the long-term correlations,
Eq.~(\ref{eq:fdtsimdth}).  To relate the exponent from
Eq.~(\ref{eq:srm0simm0b}), $\beta$, to the temporal correlation
exponent $\gamma$, from Eq.~(\ref{corr}), and therefore to $H$,
one can first rewrite Eq.~(\ref{eq:rlnm1m0}) as:
\begin{eqnarray*}
  r &=& \ln \frac{m_1}{m_0} = \ln \frac{m_0 + \Delta m}{m_0} \quad \textrm{ with } \Delta m=m_1-m_0 \\
  &=& \ln \left(\frac{\Delta m}{m_0}+1\right) 
\approx \frac{\Delta m}{m_0} \quad \textrm{ for small } \frac{\Delta m}{m_0}.
\end{eqnarray*}
Next, the total increment of messages $\Delta m$ is expressed in 
terms of smaller increments $\mu(t)$, such as messages per day: 
\begin{equation*}
\Delta m = \sum_{t=t_0+1}^{t_0+\Delta t}\mu(t)
\, ,
\end{equation*}
which is (assuming stationarity) statistically equivalent to
$\Delta m = \sum_{t=1}^{\Delta t}\mu(t)
\, ,
$
and one can write 
$r\approx \frac{1}{m_0}\sum_{t=1}^{\Delta t}\mu(t)$
for the growth rate.
The conditional average growth is then
\begin{eqnarray*}
\langle r(m_0)\rangle &=& \langle 
\frac{1}{m_0}\sum_{t=1}^{\Delta t}\mu(t) \rangle \approx 
\frac{1}{m_0}\sum_{t=1}^{\Delta t}\langle\mu(t)\rangle 
\, .
\end{eqnarray*}
Then, the conditional standard deviation $\sigma(m_0) = \sqrt{\langle
  [r(m_0)-\langle r(m_0)\rangle]^2\rangle} \, ,$ can be written in
terms of the auto-correlation function as follows:
\begin{eqnarray*}
r(m_0)-\langle r(m_0)\rangle &=& 
\frac{1}{m_0} \left(\sum_{t=1}^{\Delta t}\mu(t)-
\sum_{t=1}^{\Delta t}\langle \mu(t)\rangle\right) \\
\left[r(m_0)-\langle r(m_0)\rangle\right]^2 &=& 
\frac{1}{m_0^2} \left(\sum_{t=1}^{\Delta t}
\left(\mu(t)-\langle\mu(t)\rangle\right)\right)^2 \\
\langle\left[r(m_0)-\langle r(m_0)\rangle\right]^2\rangle &\approx& 
\frac{1}{m_0^2} \sum_i^{\Delta t}\sum_j^{\Delta t}\sigma^2_\mu C(j-i) 
\, ,
\end{eqnarray*}
where $C(\Delta t)=\frac{1}{\sigma^2_\mu}
\langle 
\left[\mu(t)-\langle\mu(t)\rangle\right]
\left[\mu(t+\Delta t)-\langle\mu(t)\rangle\right]
\rangle$ 
is the auto-correlation function of $\mu(t)$ 
and $\sigma_\mu$ is the standard deviation of $\mu(t)$. 
The auto-correlation function $C(\Delta t)$ measures the
interdependencies between the values of the record $\mu(t)$. For
uncorrelated values, $C(\Delta t)$ is zero for $\Delta t>0$, because on
average positive and negative products of the record will cancel out
each other.  In the case of short-term correlations, $C(\Delta t)$ has
a characteristic decay time, $\Delta t_\times$.  A prominent example
is the exponential decay $C(\Delta t)\sim
\exp(-\Delta t/\Delta t_\times)$.
Long-term correlations are described by a slower decay namely a
power-law,
\begin{equation}
C(\Delta t)\sim (\Delta t)^{-\gamma} 
\, ,
\label{corr}
\end{equation}
with the correlation exponent~$0<\gamma<1$ which is related to the
fluctuation exponent $H$ from Eq. (\ref{eq:fdtsimdth}) by
$\gamma=2-2H$ \cite{Feder1988}. 
We note that $\gamma=1$ (or $\gamma>1$) corresponds to an
uncorrelated record with $H=1/2$.  A key-property of long-term
correlations is a pronounced mountain-valley structure in the records
\cite{Feder1988}.
Statistically, large values of $\mu(t)$ are likely to be followed by
large values and small values by small ones.  Ideally, this holds on
all time scales, which means a sequence in daily, weekly or monthly
resolution is correlated in the same way as the original sequence.

Assuming long-term correlations asymptotically decaying as in
Eq.~(\ref{corr}), we approximate the double sum with integrals and
obtain:
\begin{eqnarray*}
  \langle\left[r(m_0)-\langle r(m_0)\rangle\right]^2\rangle &\approx& 
  \frac{1}{m_0^2}\sigma_\mu^2 
  \int\!\!\!\int_{1}^{\Delta t}\!\!(j-i)^{-\gamma}{\rm d}j{\rm d}i \sim
  \frac{1}{m_0^2}\sigma_\mu^2\left(\Delta t\right)^{2-\gamma}
  \, .
\end{eqnarray*}

In order to relate $\Delta t$ and $m_0$, one can use $\Delta t =
x\,t_0 \, , $ where $x$ is an arbitrary (small) constant, that simply
states how large $\Delta t$ is compared to $t_0$, and $ m_0\sim t_0 \,
, $ which states that the number of messages is proportional to time
assuming stationary activity.  Using these two arguments we obtain:
\begin{eqnarray*}
\langle\left[r(m_0)-\langle r(m_0)\rangle\right]^2\rangle &\approx&
\frac{1}{m_0^2}\sigma_\mu^2\left(x\right)^{2-\gamma}
\left(t_0\right)^{2-\gamma} \sim \sigma_\mu^2 m_0^{-\gamma}, \\
\sigma(m_0) &\sim& \sigma_\mu m_0^{-\gamma/2}
\, .
\end{eqnarray*}
Comparing with Eq.~(\ref{eq:srm0simm0b}), we finally obtain
$\beta=\gamma/2 \, , $ and with $\gamma=2-2H$:
\begin{equation}
\beta=1-H
\, .
\label{eq:beta1h}
\end{equation}

Equation (\ref{eq:beta1h}) is a scaling law formalizing the relation
between growth and long-term correlations in the activity and is
confirmed by our data.  For OC1 we measured $\beta_{\rm OC1}\approx
0.22$ yielding $H_{\rm OC1}\approx 0.78$ from Eq.~(\ref{eq:beta1h}),
which is in approximate agreement with the (maximum) exponent we
obtained by direct measurements for OC1 ($H=0.75\pm 0.05$ from
Fig.~\ref{fig:dfa}b). For OC2 we obtained $\beta_{\rm OC2}\approx
0.17$ and therefore $H_{\rm OC2}\approx 0.83$ through
Eq.~(\ref{eq:beta1h}) which is not too far from the (maximum) exponent
found by direct measurements for OC2 ($H= 0.88\pm 0.03$).
According to Eq.~(\ref{eq:beta1h}), the original Gibrat's law
($\beta_{\rm G}=0$) corresponds to very strong long-term correlations
with $H_{\rm G}=1$. This is the case when the activity on all time
scales exhibits equally strong correlations.  In contrast, $\beta_{\rm
  rnd}=1/2$ represents completely random activity $(H_{\rm rnd}=1/2)$,
as obtained for the randomized data in Fig.~\ref{fig:dfa}b,d.

The mathematical framework relating long-term correlations quantified
by $H$ and the growth fluctuations quantified by $\beta$ 
could be relevant to other complex systems. While the generalized
version of Gibrat's law has been reported for economic indicators
displaying $\beta\approx 0.2$
\cite{StanleyABHLMSS1996,PlerouAGMS1999,MatiaALMS2005},
the origin of this scaling law is not clear and still being
investigated.  Our results suggest that the value of $\beta$ could be
explained by the existence of long-term correlations in the activity
of the corresponding system ranging from firms and markets to social
and population dynamics.
In turn, Eq.~(\ref{eq:beta1h}) establishes a missing link between studies of 
growth processes in economic or social systems 
\cite{StanleyABHLMSS1996,PlerouAGMS1999,MatiaALMS2005} and 
studies of long-term correlations 
such as in finance and the economy \cite{MantegnaS1999}, 
Ethernet traffic \cite{LelandTWW1994}, 
human brain \cite{LinkenkaerHansenNPI2001} or 
motor activity \cite{IvanovHHSS2007}.
Our results foreshadow that systems involving other types of human
interactions such as various Internet activities, communication via cell
phones, trading activity, etc. may display similar growth and
correlation properties as found here, offering the possibility of 
explaining their dynamics in terms of the long-term persistence of the
individuals' behavior.

\section*{Growth of the degree in the underlying social network}

Communication among the members of a community represents a type
of a social interaction that defines a network, whereas a message
is sent either based on an existing relation between two members or
establishing a new one. There is considerable interest in the
origin of broad distributions of activity in social systems. Two
paradigms have been invoked for various applications in social
systems: the ``rich-get-richer'' idea used by Simon in 1955
\cite{SimonHA1955} and the models based on optimization strategies
as proposed by Mandelbrot \cite{MandelbortB1953}.  Regarding network
models, the preferential attachment (PA) model has been introduced
\cite{BarabasiA1999} to generate a type of stochastic scale-free
networks with a power-law degree distribution in the network
topology. Considering the social network of members linked when
they exchange at least one message (that has not been sent before), we
examine the dynamic of the number of outgoing links of each member
[the out-degree~$k(t)$] in analogy to Eqs.~(\ref{eq:srm0simm0b}).

We start from the empty set of nodes consisting of all the members in
the community and chronologically add a directed link between two
members when a messages is sent.  In analogy to the growth in the
number of messages~$m(t)$ of each member, we study the growth of the
members' out-degree~$k(t)$, i.e. the number of links to others.  We
define the growth rate of every member as
\begin{equation}
r_k=\ln \frac{k_1}{k_0}
\, ,
\label{eq:krgrowth}
\end{equation}
where $k_0\equiv k(t_0)$ is the out-degree of a member at time $t_0$
and $k_1\equiv k(t_1)$ is the out-degree at time $t_1$.  Again, there
is a growth rate for each member $j$, but for a better readability, we
skip the index.  In Fig.~\ref{fig:gibratkout} we study $\langle
r_k(k_0)\rangle$, the average growth rate conditional to the initial
out-degree~$k_0$, and $\sigma_k (k_0)$, the standard deviation of the
growth rate conditional to the initial out-degree~$k_0$ for OC1 and
OC2.  We obtain almost constant average growth $\langle
r_k(k_0)\rangle$ as a function of $k_0$ as in the study of messages. 

The conditional standard deviation of the network-degree, $\sigma_k
(k_0)$, is shown in Fig.~\ref{fig:gibratkout} for both social
communities.
We obtain a power-law relation analogous to Eq.~(\ref{eq:srm0simm0b}):
\begin{equation}
\sigma_k(k_0)\sim k_0^{-\beta_k}
\, ,
\label{sigma_k}
\end{equation}
with fluctuation exponents very similar to those found for the number
of messages, namely $\beta_{k,{\rm OC1}}=0.22\pm 0.02$ for OC1 and
$\beta_{k,{\rm OC2}}=0.17\pm 0.08$ for OC2.
This values are consistent with those we obtained for the activity 
of sending messages.

Next, we consider the preferential attachment model which has been
introduced to generate scale-free networks \cite{BarabasiA1999} with
power-law degree distribution~$P(k)$ of the type investigated in the
present study. Essentially, it consists of subsequently adding nodes
to the network by linking them to existing nodes which are chosen
randomly with a probability proportional to their degree.  We consider
the undirected network and study the degree growth properties using
Eqs.~(\ref{eq:krgrowth}) and (\ref{sigma_k}) and calculate the
conditional average growth rate $\langle r_{\rm PA}(k_0)\rangle$ and
the conditional standard deviation $\sigma_{\rm PA}(k_0)$.  The
times~$t_0$ and~$t_1$ are defined by the number of nodes attached to
the network.  
Figure~2 in the SI Sec.~IV 
shows the results where an average degree $\langle k\rangle =20$;
$50,000$~nodes in~$t_0$, and $100,000$ nodes in~$t_1$ were chosen.  We
find constant average growth rate that does not depend on the initial
degree $k_0$.  The conditional standard deviation is a function of
$k_0$ and exhibits a power-law decay characterized by
Eq.~(\ref{sigma_k}), respectively Eq.~(\ref{eq:srm0simm0b}), 
with $\beta_{\rm PA}=1/2$.
The value $\beta_{\rm PA}=1/2$ in Eq.~(\ref{eq:beta1h}) corresponds to
$H=1/2$ indicating complete randomness.  There is no memory in the
system.  Since each addition of a new node is completely independent
from precedent ones, there cannot be temporal correlations in the
activity of adding links. Therefore, purely preferential attachment
type of growth is not sufficient to describe the social network
dynamics found in the present study and further temporal correlations
have to be incorporated according to Eq.~(\ref{eq:fdtsimdth}). 

For the PA model it has been shown that the degree of each node
grows in time as $k(t)\sim \left(\frac{t}{t^*}\right)^b$, where
$t^*$ is the time when the corresponding node was introduced to the
system and $b=1/2$ is the dynamics exponent in growing network
models \cite{AlbertB2002}.  Accordingly, the growth rate is given by
$r_{\rm PA}=b \ln\frac{t_1}{t_0}$, which is constant independent
of $k_0$, in accordance with our numerical findings.  Furthermore,
in SI Sec.~IV we obtain analytically the exponent $\beta_{\rm PA}=1/2$ 
confirming the numerical results, as well. Interestingly,
an extension of the standard PA model has been proposed
\cite{BianconiB2001} that takes into account different fitnesses of the
nodes to acquiring links involving a distribution of $b$-exponents
and therefore a distribution of growth rates.  This model opens the
possibility to relate the distribution of fitness values to the
fluctuations in the growth rates, a point that requires further
investigation.

\section{Discussion}

From a statistical physics point of view, the finding of
long-term correlations opens the question of the origin of such a
persistence pattern in the communication.  At this point we speculate on
two possible scenarios, which require further studies.  The question
is whether the finding of an exponent $H>0.5$ is due to a power-law
(Levy type) distribution \cite{ShlesingerFK1987,BuldyrevGHPSS1993} in
the time interval between two messages of the same person or just
from pure correlations or long-term memory in the activity of people.  
In the first 
scenario, the intervals between the messages follow a power-law
\cite{BarabasiAL2005,GersteinM1964}. 
Accordingly, the activity pattern comprises
many short intervals and few long ones, implying persistent epochs
of small and large activity.  This fractal-like activity leads to
long-term correlations with $H>1/2$ (see the analogous problem of
the origin of long-term correlations in DNA sequences as discussed
in \cite{BuldyrevGHPSS1993}).  This scenario implies a direct link
between the correlations and the distribution of inter-event
intervals which can be obtained analytically.  In the second
scenario, the intervals between the messages do not follow a Levy
type distribution, but the value of the time intervals are not
independent of each other, again representing long-term persistence.
For example, the distribution of inter-event times could be stretched
exponential (see recent work on the study of extreme events of 
climatological records exhibiting long-term correlations 
\cite{BundeEKH2005}).
Thus, deciding between these two possible scenarios
for the origin of correlations in activity requires an extended analysis of
inter-event intervals as well as correlations 
to determine whether the behavior is
Levy-like or pure memory like. A careful statistical analysis
is needed which will be the focus of future research.

To some extent, the human nature of persistent interactions enables
the prediction of the actants' activity.  Our finding implies that
traditional mean-field approximations based on the assumption that the
particular type of human activity under study can be treated as a
large number of independent random events (Poisson statistics) may
result in faulty predictions.  On the contrary, from the growth
properties found here, one can estimate the probability for members of
certain activity level to send more than a given number of messages in
the future.  This result may help to improve the proper allocation of
resources in communication-based systems ranging from economic
markets to political systems.  As a byproduct, our finding that the
activity of sending messages exhibits long-term persistence suggests
the existence of an underlying long-term correlated process.
This can be understood as an unknown individual state driven by
various internal and external stimuli
\cite{HedstroemP2005,KentsisA2006} providing the probability to send
messages.  In addition, the memory in activity found here could be the
origin of the long-term persistence found in other records
representing a superposition of the individuals' behavior, such as
the Ethernet traffic \cite{LelandTWW1994}, highway traffic, stock
markets, and so forth.

\section{Materials and Methods}
\label{sec:matmathe}

\subsection*{Calculations of $\langle r(m_0)\rangle$, $\sigma(m_0)$ and 
optimal times $t_0$ and $t_1$}

The average growth rate, $\langle r(m_0) \rangle$, and the standard
deviation,
$\sigma(m_0) =\sqrt{\langle r(m_{0})^2 \rangle - \langle r(m_{0}) \rangle^{2}}$, 
are defined as follows.  Calling $P(r|m_0)$ the conditional
probability density of finding a member with growth rate $r(m_0)$ with
the condition of initial number of messages $m_0$, then we obtain:

\begin{equation}
\langle r(m_0) \rangle = \int \!\! r P(r|m_0) \, {\rm d}r
\, ,
\end{equation}
and 
\begin{equation}
\langle r(m_{0})^{2} \rangle = \int \!\! r^{2} P(r|m_0) \, {\rm d}r
\, .
\end{equation}

In order to calculate the growth rate Eq.~(\ref{eq:rlnm1m0}), one has
to choose the times~$t_0$ and~$t_1$ in the period of data
acquisition~$T$.  Naturally, it is best to use all data in order to
have optimal statistics.  Accordingly, $t_1$ is chosen best at the end
of the available data ($t_1=T$).  We argue that if the choice of $t_0$
is too small, then $m(t_0)$ is zero for many members (those that send
their messages later), which are then rejected in the calculation
because of the division in Eq.~(\ref{eq:rlnm1m0}).  Conversely, if
$t_0$ is chosen too large, then there is not enough time to observe
the member's activity and $r=0$ will occur frequently, indicating no
change (members have sent their messages before).  Thus, there must be
an optimal time in between.  
In SI Sec.~II, Fig.~1, 
we plot, as a 
function of~$t_0$, the number of members with at least one message
at~$t_0$ [$m_0>0$] and further exhibit at least some activity until
$t_1=T$ [$m_1-m_0>0$].  For both online communities we find an
optimal~$t_0$ in the middle of the period of observation $t_0=T/2$, a
value that is used for the analysis in the main text.

\subsection*{Shuffling of the message data}

The raw data comprises one entry for each message consisting of the
time when the message is sent, the sender identifier and the receiver
identifier. For example:
\begin{center}
\texttt{
\begin{tabular}{ccc}
time & sender & receiver \\
1 & a & b \\
2 & a & c \\
4 & b & a \\
6 & c & d \\
7 & a & b \\
\dots & & \\
\end{tabular}
}
\end{center}
This means, 
at $t=1$ member \texttt{a} sends a message to member \texttt{b}, 
at $t=2$ member \texttt{a} sends a message to member \texttt{c}, 
and so on. 

The randomized surrogate data set is created by randomly swapping the
instants (\texttt{time}) at which the messages are sent between two
events chosen at random.  Thus, each message entry randomly obtains
the time of another one.  This means the total number of messages is
preserved and the associations between them get shuffled.  Temporal
correlations are destroyed, but the set of instants at which the
messages are sent remains unchanged. For instance, swapping events at
$t=1$ and $t=6$ results in: $t=1$,
\texttt{c}$\,\rightarrow\,$\texttt{d}, and $t=6$,
\texttt{a}$\,\rightarrow\,$\texttt{b}.

\subsubsection*{Acknowledgments}

We thank NSF-SES-0624116 for financial support and C. Briscoe,
L. K. Gallos and H. D. Rozenfeld for discussions. 
F. L. acknowledges financial support from The
Swedish Bank Tercentenary Foundation.

\clearpage


\newpage


\setcounter{figure}{0}
\begin{figure}[h]
\includegraphics[width=0.5\textwidth]{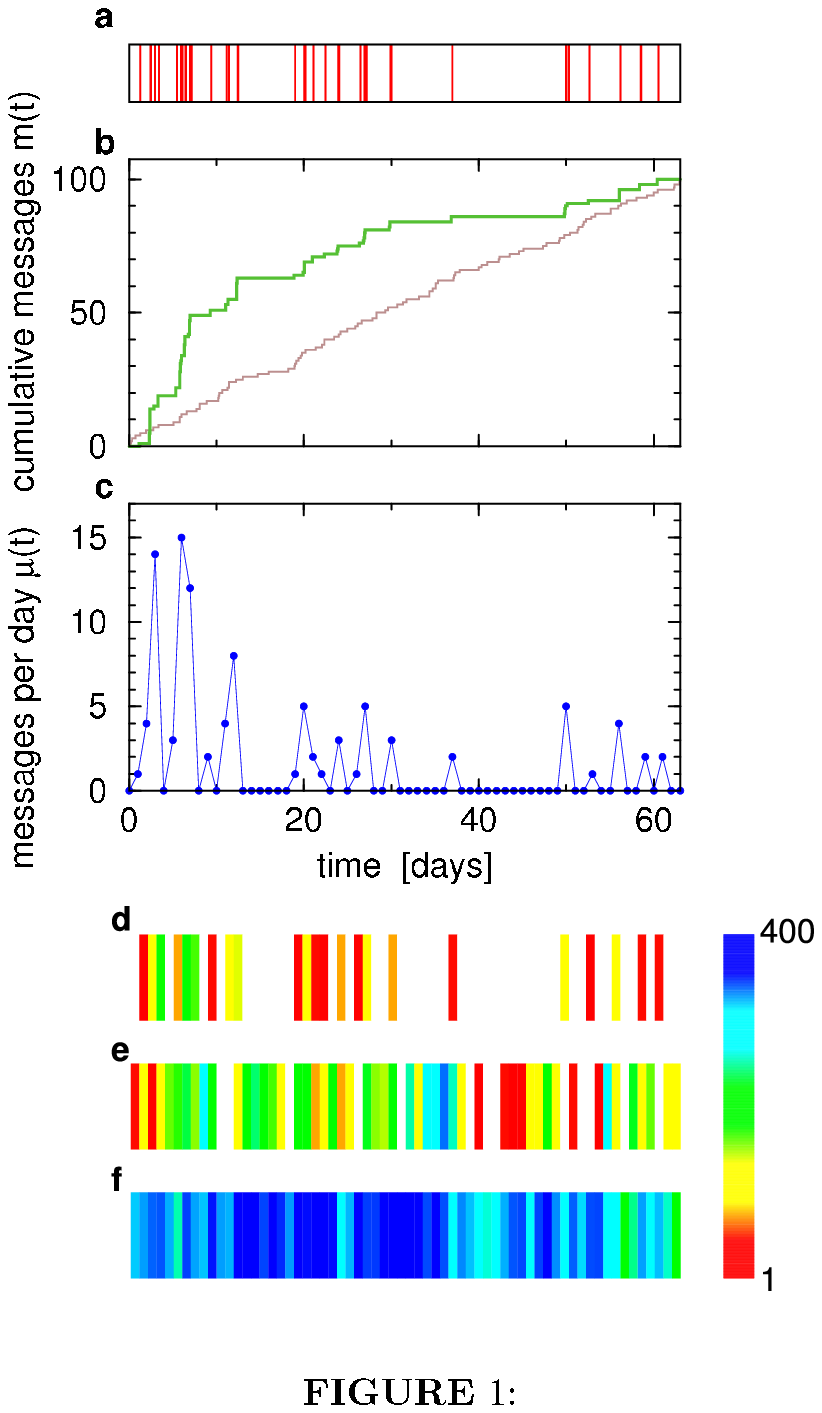}
\caption{
{\bf A typical example of an 
individuals' message activity.}
{\bf a}, Instants at which messages were sent by a member belonging to OC1. 
{\bf b}, Cumulative number of messages~$m(t)$ (green) and 
the same but with the messages placed at random (brown).
{\bf c}, Sequence of number of messages sent per day, $\mu(t)$, 
for the same individual.
{\bf d,e,f}, Color coded sequences $\mu(t)$ for members sending 
$M=100$; 1,000; or 10,000 messages overall, respectively. 
The color is proportional to the logarithm of the number of messages per day 
(red: $1$~message, blue: $400$~messages, white for no message).
}
\label{fig:example}
\end{figure}

\setcounter{figure}{1}
\begin{figure}[h]
\includegraphics[width=\textwidth]{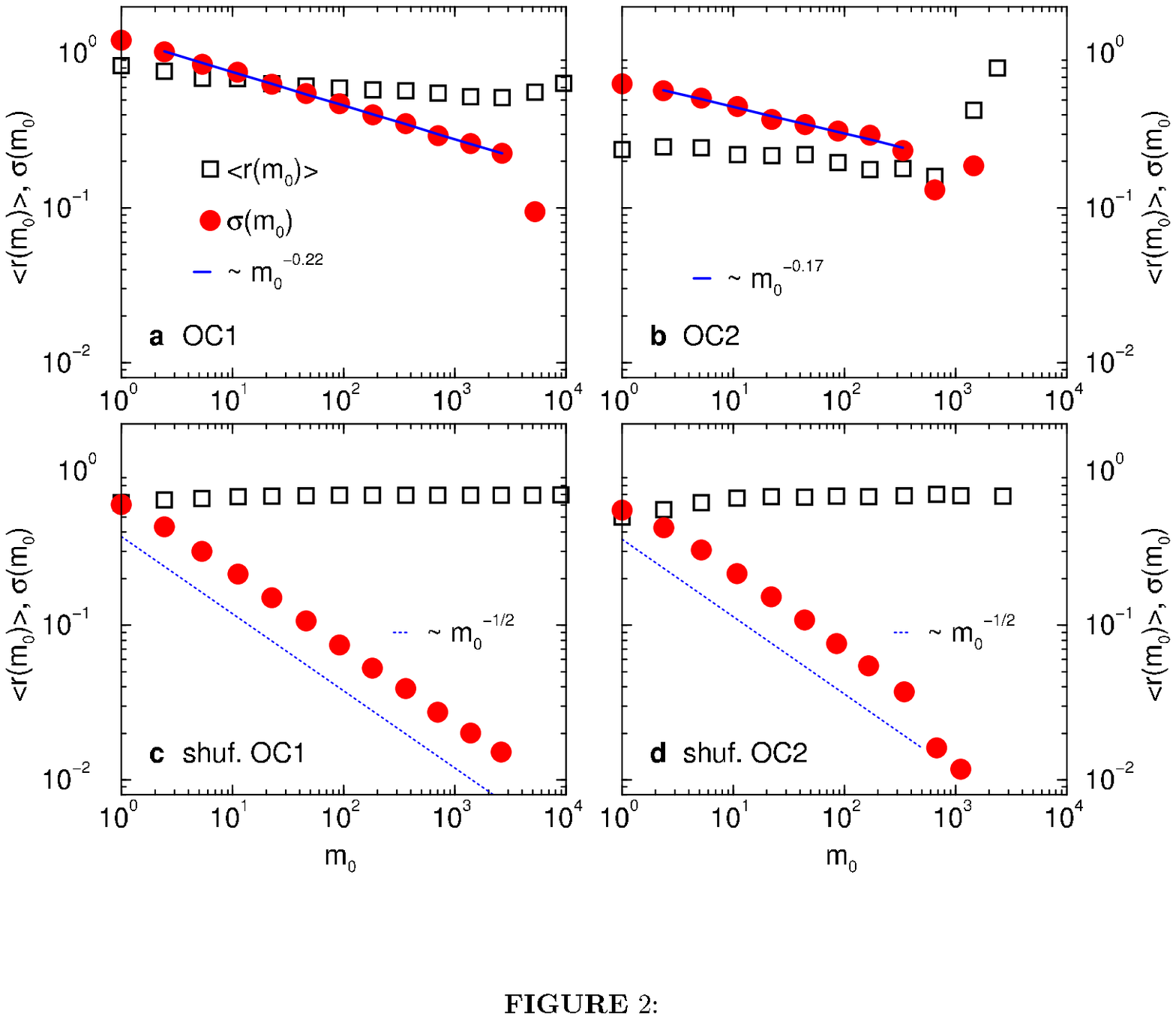}
\caption{
{\bf Average and standard deviation of
the growth rate versus number of messages.} 
{\bf a}, Results for OC1. 
The average growth rate of messages conditional to $m_0$ is
almost constant and the standard deviation decays with an exponent
$\beta_{\rm OC1}= 0.22\pm 0.01$. 
{\bf b}, Results for OC2. 
The standard deviation conditional to $m_0$ decays with an exponent
$\beta_{\rm OC2}= 0.17\pm 0.03$. 
{\bf c}, Results for OC1, when the messages are shuffled, 
displaying $\beta_{\rm rnd}=1/2$. 
{\bf d}, Results for OC2, when the messages are shuffled. 
In all cases $t_0$ corresponds to half of the period of data acquisition 
and $t_1$ to the end, which we found to provide optimal statistics 
(see SI Fig.~1).
}
\label{fig:gibrat}
\end{figure}

\setcounter{figure}{2}
\begin{figure}[h]
\includegraphics[width=\textwidth]{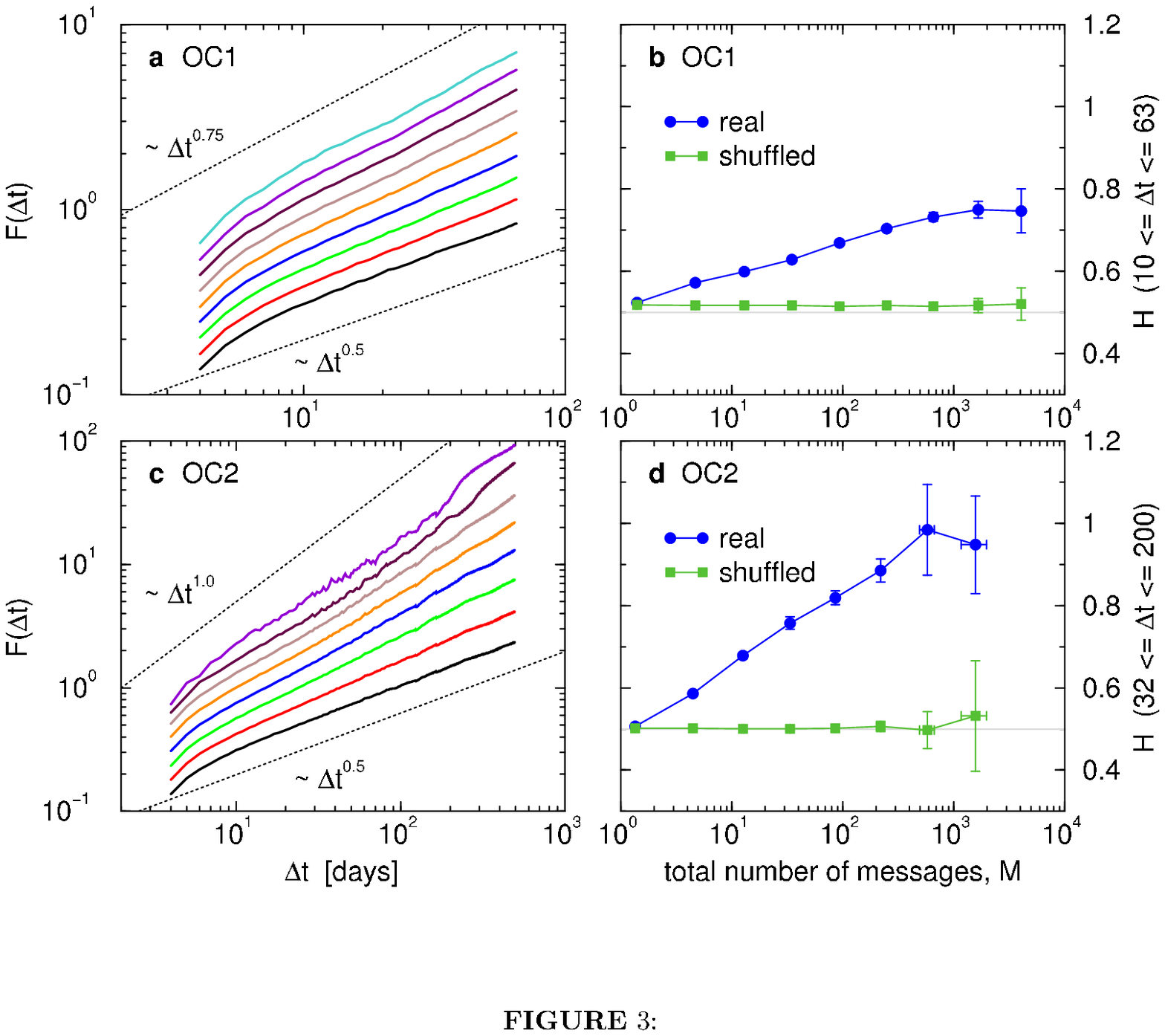}
\caption{
{\bf Long-term correlations in the
message activity of OC1 (a and b) and OC2 (c and d).} 
{\bf a}, DFA fluctuation functions averaged conditional to $M$, 
the total number of messages sent by each member 
(black: 1-2, red: 3-7, green: 8-20, blue: 21-54, 
orange: 55-148, brown: 149-403, maroon: 404-1096,
violet: 1097-2980, turquoise: 2981-8103). 
The dotted lines serve as guides, 
the one in the bottom corresponds to the uncorrelated case,
while the one in the top corresponds to the exponent $0.75$. 
{\bf b}, Fluctuation exponent $H$ measured from panel a on the scales
$10\textrm{ days}\le \Delta t \le 63\textrm{ days}$ as a function of
the total number of messages sent, $M$, for real (blue) and
individually shuffled (green) records. 
{\bf c}, DFA fluctuation functions averaged conditional to $M$ 
[colors as in (A)]. 
The dotted lines correspond to the uncorrelated case (bottom) 
and to the exponent $1$ (top). 
{\bf d}, Fluctuation exponents obtained from panel c on the scales 
$32\textrm{ days}\le \Delta t \le 200\textrm{ days}$ as a function of 
the total number of messages sent, $M$. 
Due to weak statistics causing large error bars we do not consider the
last two values for $M>500$ as reliable. 
For clarity the fluctuation functions in panels a and c are shifted vertically.
}
\label{fig:dfa}
\end{figure}

\setcounter{figure}{3}
\begin{figure}[h]
\includegraphics[width=\textwidth]{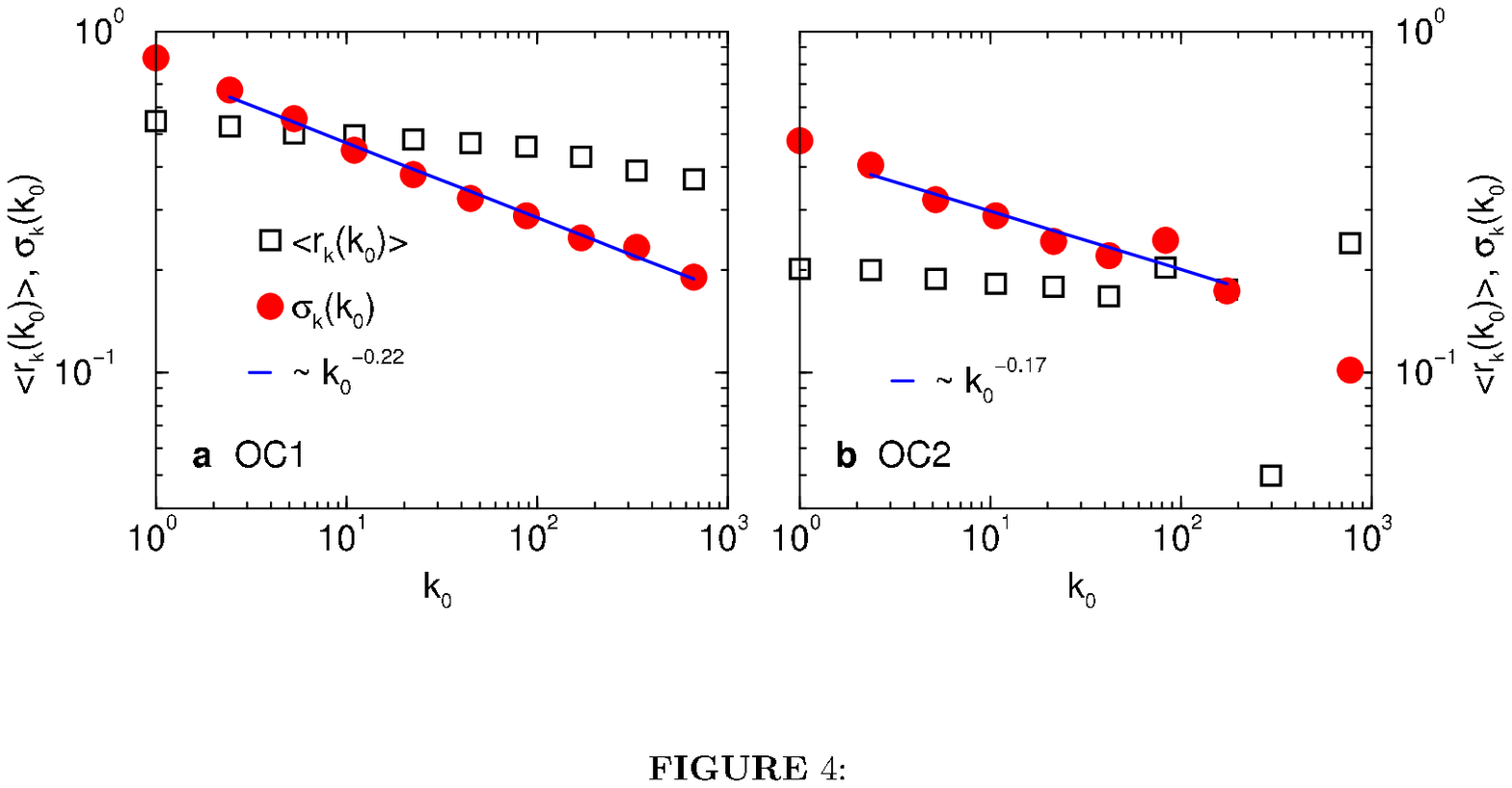}
\caption{
{\bf Mean out-degree growth rate
and standard deviation versus initial out-degree.}
{\bf a}, Results for OC1. 
The average growth of out-degree conditional to the out-degree at $t_0$ 
is almost constant. 
The standard deviation decays with an exponent 
$\beta_{k,{\rm OC1}}= 0.22\pm 0.02$. 
{\bf b}, Results for OC2. 
The standard deviation conditional to the out-degree at $t_0$ 
decays with an exponent $\beta_{k,{\rm OC2}}= 0.17\pm 0.08$. 
The quantities are analogous to those of Fig.~\ref{fig:gibrat} except that 
here the growth rate of the out-degree $r_k$ is considered 
instead of the number of messages sent.
}
\label{fig:gibratkout}
\end{figure}

\clearpage

\newpage

%
%

\setcounter{section}{0}

\centerline{\bf SUPPORTING INFORMATION (SI)}

\vspace{0.5cm}

\begin{center}

{\bf Scaling laws of human interaction activity}

\vspace{0.5cm}

Diego Rybski, 
Sergey V. Buldyrev,
Shlomo Havlin,\\
Fredrik Liljeros, 
and Hern\'an A. Makse

\end{center}

\vspace{0.5cm}

\section{Notation}
\label{ssec:notation}

\begin{enumerate}
\item Member $j$ sends his/her $n$th message at time $t_j(n)$, where
  $1\le n\le M_j$ and $M_j$ is the total number of messages sent by
  $j$ in the time of data acquisition $T$. The sequence of counts
  defined as the number of messages in the period $\delta t$, is given
  by
\begin{equation}
\mu_j^{\delta t}(t)=\sum_{n,t_j(n)\in[t,t+\delta t]} a_j(n)
\, ,
\end{equation}
where $a_j(n)=1$. 
In addition, the periods are non-overlapping, $t=i\delta t$ 
with integer $i$, and therefore $1\le t_j(n) \le T$. 
In the case of daily resolution $\delta t=1$ day.

\item The cumulative number of messages that a member sends until time
  $t$ is:

\begin{equation}
m_j^{\delta t}(t)=\sum_{t'=1}^{t}\mu_j^{\delta t}(t')
\, .
\end{equation}
In particular, $m_j(1)=\mu_j(1)$ and $m_j(T)=M_j$.

\item The displacement of the random walk is the cumulative sum of the
  normalized $\mu_j^{\delta t}(t)$:
\begin{equation}
  Y_j^{\delta t}(t) = 
  \sum_{t'=1}^{t}(\mu_j^{\delta t}(t')-\langle\mu_j^{\delta t}(t)\rangle)
  \, ,
\end{equation}
where $\langle\mu_j^{\delta t}(t)\rangle$ is the average of 
$\mu_j^{\delta t}(t)$ in time $t$. 
The root-mean-square displacement after $\Delta t$ is defined as
\begin{equation}
F_j^{\delta t}(\Delta t) = 
\sqrt{\langle[Y_j^{\delta t}(t+\Delta t)-Y_j^{\delta t}(t)]^2\rangle}_t
\, ,
\end{equation}
where the average is performed over the time $t$. 
Additionally, we
perform an average over members $j$ with activity level $M$ 
and define
\begin{equation}
  (F^{\delta t}(\Delta t))^2_{M} = 
  \langle (F_j^{\delta t})^2|M\rangle_j
  \, .
\end{equation}

\item
For simplicity, in the main text we skip the index $j$ as well as $\delta t$ 
and write $\mu(t)$, $m(t)$, $Y(t)$, as well as $F(\Delta t)$.

\item To investigate the growth in the number of messages we use the
  quantities $r=\ln\frac{m_1}{m_0}$, $\langle r(m_0)\rangle$,
  $\sigma(m_0)$ and the exponents $\beta_{\rm OC1}$, $\beta_{\rm
    OC2}$, $\beta_{\rm G}$, $\beta_{\rm rnd}$.

\item
To investigate the growth of the degree we use the quantities
$r_k=\ln\frac{k_1}{k_0}$, 
$\langle r_k(k_0)\rangle$, 
$\sigma_k(k_0)$ 
and the exponents
$\beta_{k, {\rm OC1}}$; $\beta_{k, {\rm OC2}}$.

\item
For the growth of the degree in the preferential attachment model 
we use the quantities
$r_{\rm PA}=\ln\frac{k_1}{k_0}$, 
$\langle r_{\rm PA}(k_0)\rangle$, 
$\sigma_{\rm PA}(k_0)$ 
and the exponent 
$\beta_{\rm PA}$.

\end{enumerate}

\section{Optimal times $t_0$ and $t_1$}
\label{ssec:opttime}

\begin{figure}
\begin{centering}
\includegraphics[width=\textwidth]{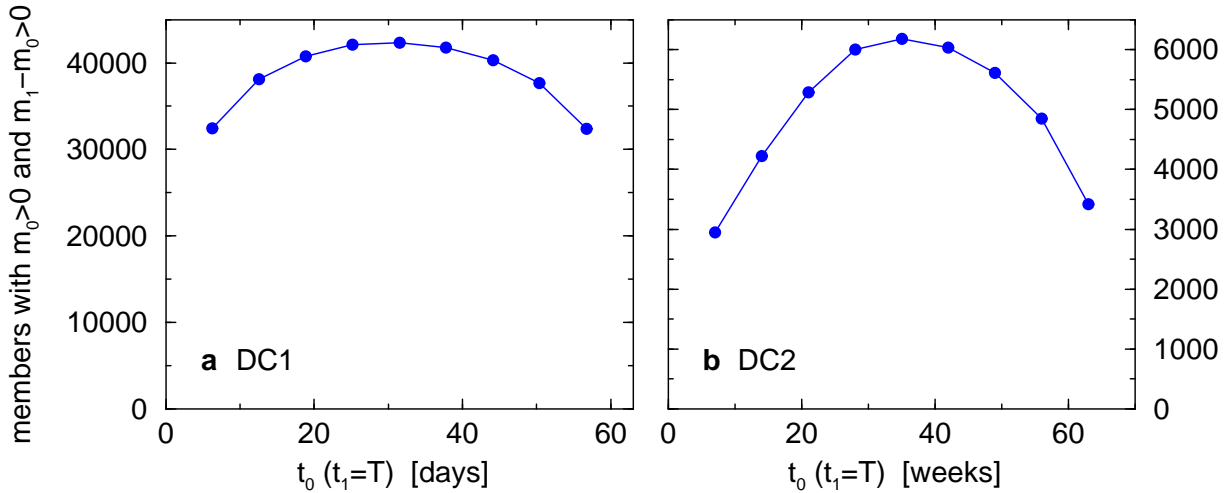}
\caption{
Optimal times~$t_0$ and~$t_1$. 
The panels show for {\bf a}, OC1, and {\bf b}, OC2, the number of members with 
both, $m_0>0$ and $m_1-m_0>0$. 
While $t_1$ obviously is optimal at the end of the period, 
$t_0$ is varied to find the value for which the number of members -- 
with at least one message until $t_0$ and at least one new message 
between $t_0$ and $t_1$ -- 
is maximal. 
\label{fig:optime}
}
\end{centering}
\end{figure}

Figure ~\ref{fig:optime} displays the optimal times~$t_0$ and~$t_1$ to
calculate the growth rates for OC1 (panel~a) and OC2 (panel~b).  

\section{Details on the quantification of long-term correlations using 
Detrended Fluctuation Analysis}
\label{ssec:ltcdfa}

Statistical dependencies between the values of a record $\mu(t)$ with
$t=1, \dots, T$ can be characterized by the auto-correlation function
\begin{equation}
C(\Delta t) = \frac{1}{\sigma_\mu^2(T-\Delta t)}
\sum_{t=1}^{T-\Delta t}
\left[\mu(t)-\langle\mu(t)\rangle\right]
\left[\mu(t+\Delta t)-\langle\mu(t)\rangle\right]
\, ,
\end{equation}
where $T$ is the length of the record $\mu(t)$,  
$\langle\mu(t)\rangle$ its average, and $\sigma_\mu$ its 
standard deviation.
For {\it uncorrelated} values of $\mu(t)$, 
$C(\Delta t)$ is zero for $\Delta t>0$, 
because on average positive and negative products will cancel each 
other out.
In the case of {\it short-term correlations} $C(\Delta t)$ 
has a characteristic decay time~$\Delta t_\times$. 
A prominent example is the exponential decay 
$C(\Delta t)\sim \exp(-\Delta t/\Delta t_\times)$. 
{\it Long-term correlations} are 
described by a slower decay, e.g. diverging $\Delta t_\times$,  
namely a power-law, 
\begin{equation}
C(\Delta t)\sim (\Delta t)^{-\gamma}
\label{eq:Cssgamma}
\, , 
\end{equation}
with the correlation exponent~$0<\gamma<1$.

Detrended Fluctuation Analysis (DFA) is a well studied method to 
quantify long-term correlations in the presence of non-stationarities 
\cite{PengBHSSG94}. 
The analysis of a considered record $\mu(t)$ of length~$T$ consists 
of \ref{it:dfaff}~steps: 

\begin{enumerate}

\item
Calculate the cumulative sum, the so-called profile:
\begin{equation}
Y(t)=\sum_{t'=1}^{t}\left(\mu(t')-\langle\mu(t)\rangle\right)
\label{eq:dfaprofile}
\, .
\end{equation}

\item
Separate the profile~$Y(t)$ into $T_{\Delta t}={\rm int}\frac{T}{\Delta t}$ 
segments of length~$\Delta t$. 
Often, the length of the record is not a multiple of~$\Delta t$. 
In order not to disregard information, the segmentation procedure is 
repeated starting from the end of the record and one obtains 
$2T_{\Delta t}$~segments.

\item
\label{it:dfadetrend}
Locally detrend each segment~$\nu$ by determining best polynomial 
fits $p^{(n)}_\nu(t)$ of order~$n$ and subsequently subtract it from the profile: 
\begin{equation}
Y_{\Delta t}(t) = Y(t)-p^{(n)}_\nu(t)
\, .
\end{equation}

\item 
Calculate for each segment the variance (squared residuals) of the 
detrended~$Y_{\Delta t}(t)$ 
\begin{equation}
F_{\Delta t}^2(\nu) = 
\frac{1}{\Delta t}\sum_{j=1}^{\Delta t}\left(Y_{\Delta t}^2
\left[(\nu-1)\Delta t+j\right]\right) 
\label{eq:Fs21ssum}
\end{equation}
by averaging over all values in the corresponding 
$\nu$th segment.

\item
\label{it:dfaff}
The DFA fluctuation function is given by the square-root of the average over 
all segments:
\begin{equation}
F(\Delta t) = 
\left[\frac{1}{2T_{\Delta t}}\sum_{\nu=1}^{2T_{\Delta t}}
F_{\Delta t}^2(\nu)\right]^{1/2}
\, .
\label{F}
\end{equation}
The averaging of $F_{\Delta t}^2(\nu)$ is additionally performed 
over members of similar activity level $M$.
\end{enumerate}

If the record $\mu(t)$ is long-term correlated according to a power-law 
decaying auto-correlation function, Eq.~(\ref{eq:Cssgamma}), 
then $F(\Delta t)$ increases for large scales~$\Delta t$ also as a power-law:
\begin{equation}
F(\Delta t) \sim (\Delta t)^H
\, ,
\label{eq:Fssimsh}
\end{equation}
where the fluctuation exponent~$H$ is analogous to the well-known 
Hurst exponent \cite{Feder1988}.
The exponents are related via  
\begin{equation}
H=1-\gamma/2 \enspace , \qquad \gamma=2-2H
\, .
\end{equation}
When $\gamma=1$ then $H_{\rm rnd}=1/2$, that is the case of
uncorrelated dynamics.  If the correlations decay faster than
$\gamma>1$ then the random exponent $H_{\rm rnd} =1/2$ is still
recovered.  Long-term correlations imply $0<\gamma<1$ and $1/2<H<1$.
In practice, one plots $F(\Delta t)$ versus $\Delta t$ in
double-logarithmic representation, determines the exponent~$H$ on
large scales and quantifies the correlation exponent~$\gamma$.  The
order of the polynomials $p^{(n)}_\nu$ determines the detrending
technique which is named DFA$n$, DFA$0$ for constant detrend, DFA$1$
for linear, DFA$2$ for parabolic, etc.

The subtraction of the average in Eq.~(\ref{eq:dfaprofile}) is only
necessary for DFA$0$.  By definition the corresponding fluctuation
function is only given for $\Delta t\ge n+2$.  The detrending order
determines the capability of detrending.
Since the local trends are subtracted from the profile, only trends of
order~$n-1$ are subtracted from the original record~$\mu(t)$.
Throughout the paper we show the results using DFA$2$ which we found
to be sufficient in terms of detrending.

Since the fluctuation functions~$F(\Delta t)$ for single users are
very noisy, it is useful to average fluctuation functions among
various members.  Thus, we first group the members in logarithmic bins
according to their activity level, the total number of messages $M$
sent. Namely, we group all members that send 1-2, 3-7, 8-20, \dots
messages in the period of data acquisition by using bins determined by
$b={\rm int}\left(\ln M\right)$.  Next we average the fluctuation function 
among all members from each group $b$ and obtain for every activity
level of the members one DFA fluctuation function.  The error bars in
Fig.~3a,c 
of the main text were obtained by
subdividing each group and determining the standard deviations of the
fluctuation exponents from different groups of the same activity
level.

\section{Growth in the degree}
\label{ba}

\begin{figure}
\begin{centering}
\includegraphics[width=0.55\textwidth]{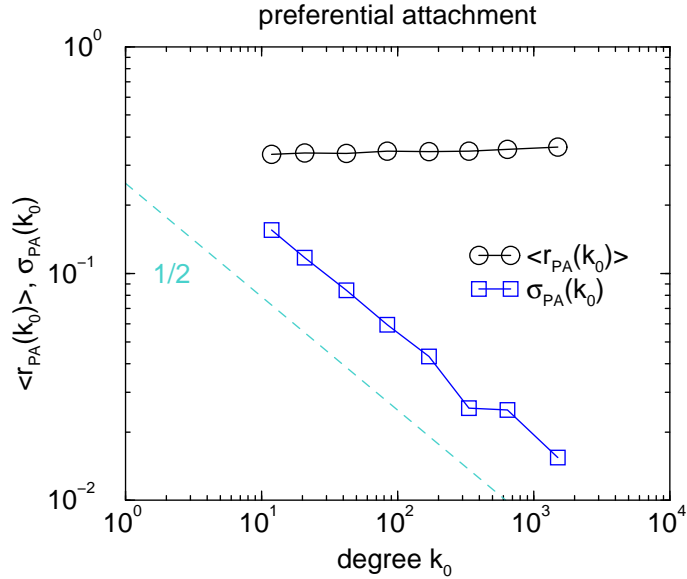}
\caption{
\label{fig:bamodel}
Growth properties of the preferential attachment model \cite{BarabasiA1999} 
discussed in the main text.
We plot the average (black circles) and standard deviation (blue squares) 
of the growth rate~$r_{\rm PA}$ conditional to $k_0$, the degree of the 
corresponding nodes at the first stage. 
}
\end{centering}
\end{figure}

Figure~\ref{fig:bamodel} shows the results of the average growth rates and
fluctuations of the growth rates as a function of the initial degree
for the preferential attachment model \cite{BarabasiA1999}.  We find a
constant average growth rate and a standard deviation decreasing as a
power law with exponent $\beta_{\rm PA}=1/2$ in 
Eq.~(7) 
in the main text.

The PA network model has been described analytically. 
In particular, it has 
been shown that each nodes' degree increases as 
\begin{equation}
k(t)\sim \left(\frac{t}{t^*}\right)^b
\, ,
\label{eq:ktpa}
\end{equation}
where $t^*$ is the time when the corresponding node was introduced to the 
system and $b$ is the dynamics exponent in growing network models 
($b=1/2$ for the standard PA) 
\cite{AlbertB2002}.
Accordingly, here the growth rate, Eq.~(6) in the main text, is 
$r_{\rm PA}=\frac{1}{2}\ln\frac{t_1}{t_0}$, which we also find in 
Fig.~\ref{fig:bamodel}.

To obtain $\sigma_{\rm PA}(k_0)$ one can use analogous considerations as for 
$\sigma(m_0)$ in the main text. 
Due to Eq.~(6) in the main text, here we have 
\begin{equation}
r_{\rm PA}\approx \frac{1}{k_0}\sum_{t=1}^{\Delta t}\kappa(t)
\, ,
\end{equation}
where $\kappa(t)$ are small increments analogous to $\mu(t)$, 
whereas Eq.~(\ref{eq:ktpa}) implies
\begin{equation}
\kappa(t)\sim (\Delta t)^{-1/2}
\, .
\end{equation}
As before, the conditional standard deviation of the growth rate is
\begin{equation}
\langle\left[r_{\rm PA}(k_0)-\langle r_{\rm PA}(k_0)\rangle\right]^2\rangle 
\approx 
\frac{1}{k_0^2} \sum_i^{\Delta t}\sum_j^{\Delta t}\sigma^2_\kappa C(j-i)
\, .
\end{equation}
In the uncorrelated case $C(j-i)=\delta_{ij}$, 
the double sum can be reduced to a single one:
\begin{equation}
\sigma^2_{\rm PA}(k_0)
= 
\frac{1}{k_0^2} \sum_i^{\Delta t}\sigma^2_\kappa(i)
\, .
\end{equation}
As shown below, $\sigma_\kappa(i) \sim i^{-1/4}$, and integration leads to 
\begin{eqnarray}
\sigma^2_{\rm PA}(k_0)
&\sim &
\frac{1}{k_0^2}\int^{\Delta t} \!\!\! i^{-1/2}{\rm d}i \\
&\sim &
\frac{1}{k_0^2} (\Delta t)^{1/2}
\, .
\end{eqnarray}
Eliminating $\Delta t$ using $k\sim t^{-1/2}$, Eq.~(\ref{eq:ktpa}), 
one obtains
\begin{equation}
\sigma_{\rm PA}(k_0) \sim k_0^{-1/2}
\, .
\end{equation}
That is, we obtain $\beta_{\rm PA}=1/2$ as found numerically.

Remains to show $\sigma_\kappa(t) \sim t^{-1/4}$.
We assume new links are set according to a Poisson process, 
whereas every new link of a node represents an event. 
The intervals between these events (asymptotically) follow an 
exponential distribution $p(\tau)=\lambda {\rm e}^{-\lambda\tau}$.
Accordingly, $\kappa(t)$ is a sequence of zeros and only one when 
a new link is set to the corresponding node. 
The standard deviation of this sequence is
\begin{equation}
\sigma_\kappa \sim \lambda^{1/2}
\, .
\end{equation}
Due to Eq.~(\ref{eq:ktpa}) the rate parameter decreases like 
\begin{equation}
\lambda(t) \sim t^{-1/2}
\, .
\end{equation}
Accordingly, 
\begin{equation}
\sigma_\kappa(t) \sim t^{-1/4}
\, .
\end{equation}

In order to extend the standard PA model, a fitness model has been 
introduced \cite{BianconiB2001} taking into account different fitnesses of 
the nodes of acquiring links and therefore involving a distribution 
of $b$-exponents.
The spread of growth rates~$r$ could be related to the 
distribution of fitness. 
On the other hand, the growth according to Eq.~(\ref{eq:ktpa}) is superimposed 
with random fluctuations that we characterize with the exponent~$\beta$.


\begin{thebibliography}{10}

\bibitem{MertonR1936}
Merton RK
\newblock (1936) 
The Unanticipated Consequences of Purposive Social Action.
{\em Am Sociol Rev} 1:894--904.

\bibitem{WeberM1968}
Weber M
\newblock (1968) {\em Economy and Society, Vol.1}.
\newblock (University of California Press, Berkley).

\bibitem{GiddensA1993}
Giddens A
\newblock (1993) {\em New Rules of Sociological Method}.
\newblock (Stanford University Press, Stanford).

\bibitem{DurkheimE1997}
Durkheim E
\newblock (1997) {\em Suicide, reprint from 1897}.
\newblock (The Free Press, New York).

\bibitem{ParetoV1896}
Pareto V
\newblock (1896) {\em Cours d'Economie Politique}.
\newblock (Droz, Geneva).

\bibitem{ZipfG1932}
Zipf G
\newblock (1932) {\em Selective Studies and the Principle of Relative Frequency
  in Language}.
\newblock (Harvard University Press, Cambridge, MA).

\bibitem{GibratR1931}
Gibrat R
\newblock (1931) {\em Les in\'egalit\'es \'economiques}.
\newblock (Libraire du Recueil Sierey, Paris).

\bibitem{SuttonJ1997}
Sutton J
\newblock (1997) 
Gibrat's Legacy.
{\em J Econ Lit} 35:40--59.

\bibitem{GabaixX1999}
Gabaix X
\newblock (1999)
Zipf's law for cities: An explanation.
{\em Q J Econ} 114:739--767.

\bibitem{HolmeEL2004}
Holme P, Edling CR, Liljeros F
\newblock (2004)
Structure and time evolution of an Internet dating community.
{\em Soc Networks} 26:155--174.

\bibitem{PaxsonF1995}
Paxson V, Floyd S
\newblock (1995) 
Wide area traffic: the failure of Poisson modeling.
{\em IEEE/ACM Trans Networking} 3:226--244.

\bibitem{DewesWF2003}
Dewes C, Wichmann A, Feldman A
\newblock (2003) {\em Proc. 2003 ACM SIGCOMM Conf. Internet Measurement
  (IMC-03)}.
\newblock (ACM Press, New York).

\bibitem{BarabasiAL2005}
Barab{\'a}si A-L
\newblock (2005) 
The origin of bursts and heavy tails in human dynamics.
{\em Nature} 435:207--211.

\bibitem{OliveiraB2005}
Oliveira JG, Barab{\'a}si A-L
\newblock (2005)
Darwin and Einstein correspondence patterns.
{\em Nature} 437:1251.

\bibitem{ZhouKKWH2008}
Zhou T, Kiet HAT, Kim BJ, Wang B-H, Holme P 
\newblock (2008)
Role of activity in human dynamics. 
{\em Europhys Lett} 82:28002.

\bibitem{StanleyABHLMSS1996}
Stanley MHR, {\it et al.}
\newblock (1996)
Scaling behaviour in the growth of companies.
{\em Nature} 379:804--806.

\bibitem{PlerouAGMS1999}
Plerou V, Amaral LAN, Gopikrishnan P, Meyer M, Stanley HE 
\newblock (1999)
Similarities between the growth dynamics of 
university research and of competitive economic 
activities.
{\em Nature} 400:433--437.

\bibitem{MatiaALMS2005}
Matia K, Amaral LAN, Luwel M, Moed HF, Stanley HE 
\newblock (2005)
Scaling Phenomena in the Growth Dynamics of Scientific Output.
{\em J Am Soc Inf Sci Tec} 56:893--902.

\bibitem{RozenfeldRABSM2008}
Rozenfeld HD, {\it et al.}
\newblock (2008)
Laws of Population Growth. 
{\em Proc Nat Acad Sci USA} 105:18702--18707.

\bibitem{Feder1988}
Feder J
\newblock (1988) {\em Fractals}, Physics of Solids and Liquids.
\newblock (Plenum Press, New York).

\bibitem{PengBHSSG94}
Peng C-K, {\it et al.}
\newblock (1994) 
Mosaic organization of DNA nucleotides. 
{\em Phys Rev E} 49:1685--1689.

\bibitem{EislerBK2008}
Eisler Z, Bartos I, Kert\'esz J
\newblock (2008) 
Fluctuation scaling in complex systems: Taylor's law and beyond.
{\em Adv Phys} 57:89--142.

\bibitem{MantegnaS1999}
Mantegna RN, Stanley HE 
\newblock (1999) {\em An Introduction to Econophysics: Correlations and
  Complexity in Finance}.
\newblock (Cambridge University Press, Cambridge).

\bibitem{LelandTWW1994}
Leland WE, Taqqu MS, Willinger W, Wilson DV 
\newblock (1994) 
On the Self-Similar Nature of Ethernet Traffic (Extended Version)
{\em IEEE/ACM Trans Networking} 2:1--15.

\bibitem{LinkenkaerHansenNPI2001}
Linkenkaer-Hansen K, Nikouline VV, Palva JM, Ilmoniemi RJ 
\newblock (2001) 
Long-range temporal correlations and scaling behavior in 
human brain oscillations.
{\em J Neurosci} 21:1370--1377.

\bibitem{IvanovHHSS2007}
Ivanov PC, Hu K, Hilton MF, Shea SA, Stanley HE 
\newblock (2007) 
Endogenous circadian rhythm in human motor activity 
uncoupled from circadian influences on cardiac 
dynamics.
{\em Proc Nat Acad Sci USA} 104:20702--20707.

\bibitem{SimonHA1955}
Simon HA 
\newblock (1955) 
On a Class of Skew Distribution Functions.
{\em Biometrika} 42:425--440.

\bibitem{MandelbortB1953}
Mandelbrot B
\newblock (1953) {\em An informational theory of the statistical structure of
  language}, ed.{} Jackson, W.
\newblock (Butterworth, London), pp. 486--504.

\bibitem{BarabasiA1999}
Barab\'asi A-L, Albert R 
\newblock (1999) 
Emergence of scaling in random networks. 
{\em Science} 286:509--512.

\bibitem{AlbertB2002}
Albert R, Barab\'asi A-L
\newblock (2002) 
Statistical mechanics of complex networks. 
{\em Rev Mod Phys} 74:47--97.

\bibitem{BianconiB2001}
Bianconi G, Barab\'asi, A-L
\newblock (2001)
Competition and multiscaling in evolving networks.
{\em Europhys Lett} 54:436-442.

\bibitem{ShlesingerFK1987}
Shlesinger MF, West BJ, Klafter J
\newblock (1987) 
L\'evy dynamics of enhanced diffusion: Application to turbulence.
{\em Phys Rev Lett} 58:1100--1103.

\bibitem{BuldyrevGHPSS1993}
Buldyrev SV, Goldberger AL, Havlin S, Peng C-K, Simons M, Stanley HE 
\newblock (1993)
Generalized L\'evy-walk model for DNA nucleotide sequences.
{\em Phys Rev E} 47:4514--4523.

\bibitem{GersteinM1964}
Gerstein GL, Mandelbrot B
\newblock (1964)
Random walk models for spike activity of single neuron.
{\em Biophys J} 4:41--68.

\bibitem{BundeEKH2005}
Bunde A, Eichner JF, Kantelhardt JW, Havlin S
\newblock (2005)
Long-Term Memory: A Natural Mechanism for the Clustering of 
Extreme Events and Anomalous Residual Times in 
Climate Records.
{\em Phys Rev Lett} 94:048701.

\bibitem{HedstroemP2005}
Hedstr\"om P
\newblock (2005) {\em Dissecting the Social: On the Principles of Analytical
  Sociology}.
\newblock (Cambridge University Press, Cambridge).

\bibitem{KentsisA2006}
Kentsis A
\newblock (2006) 
Mechanisms and models of human dynamics. 
{\em Nature} 441:E5--E6.


\end{thebibliography}
\end{document}